\documentclass[reqno]{amsart}

\usepackage[active]{srcltx}
\usepackage{amsmath,color}
\usepackage{amssymb,amsxtra,amscd}
\DeclareMathOperator{\tr}{tr}

\usepackage{amsfonts} 
\begin{document} 
\newcommand{\R}{\mathbb R} 
\newcommand{\n}{\nabla} 
\newcommand{\pa}{\partial} 
\newcommand{\eps}{\varepsilon} 
\newcommand{\di}{\mbox{\rm div}\,} 
\newcommand{\dn}{\cdot \nabla}            
\newcommand{\nd}{\nabla \cdot} 
\renewcommand{\a}{\vspace{3mm}\\} 
\renewcommand{\b}{\begin} 
\newcommand{\e}{\end} 
\newcommand{\ba}{\begin{eqnarray}} 
\newcommand{\ea}{\end{eqnarray}} 
\newcommand{\bv}{\mathbf v} 
\newcommand{\bw}{\mathbf w} 
\newcommand{\bx}{\mathbf x} 
\newcommand{\f}{\mathbf f} 
\newcommand{\bg}{\mathbf g} 
\newcommand{\bh}{\mathbf h} 
\newcommand{\bn}{\mathbf n} 
\newcommand{\bz}{\mathbf z} 
\newcommand{\bT}{\mathbf T} 
\newcommand{\bL}{\mathbf L} 
\newcommand{\bb}{\mathbf b} 
\newcommand{\bq}{\mathbf q} 
\newcommand{\bD}{\mathbf D} 
\newcommand{\bphi}{\boldsymbol \varphi}
\newcommand{\bI}{\mathbf I} 
\renewcommand{\Re}{\mathrm{Re}} 
\renewcommand{\Pr}{\mathrm{Pr}\,} 
\newcommand{\Gr}{\mathrm{Gr}\,} 
\newcommand{\Ra}{\mathrm{R}\,} 
\newcommand{\Di}{\mathrm{Di}\,} 
\newcommand{\rose}{R\r u\v zi\v cka} 
\newcommand{\ov}[1]{\overline{#1}}
\newcommand{\ovp}{\overline{p}}
\newcommand{\ovt}{\overline{\theta}}
\newcommand{\phiab}{\phi^{A,B}}
\renewcommand\thesection{\arabic{section}}
\renewcommand\theequation{\thesection.\arabic{equation}}
\newtheorem{example}[equation]{Example}
\newtheorem{remark}[equation]{Remark}

\providecommand{\comment}[1]{\vskip.3cm
\fbox{%
\parbox{0.93\linewidth}{\footnotesize #1}}
\vskip.3cm}

\providecommand{\abstmp}[2]{{#1\lvert{#2}#1\rvert}}
\providecommand{\abs}[1]{\abstmp{}{#1}}
\providecommand{\bigabs}[1]{\abstmp{\big}{#1}}
\providecommand{\Bigabs}[1]{\abstmp{\Big}{#1}}
\providecommand{\biggabs}[1]{\abstmp{\bigg}{#1}}
\providecommand{\Biggabs}[1]{\abstmp{\Bigg}{#1}}

\title{The Oberbeck--Boussinesq Approximation\\ as a Constitutive Limit}
\author{Yoshiyuki Kagei} 
\address{Graduate School of Mathematics, Kyushu University 36, Fukuoka 812-8581, Japan }

\author{Michael \rose} 
\address{Institute of Applied Mathematics, University of Freiburg,
  Eckerstr.  1, D-79104 Freiburg, Germany} 
\email{rose@mathematik.uni-freiburg.de} 
\begin{abstract} 
  \hspace*{-1em}{\bf :} We derive the usual Oberbeck--Boussinesq
  approximation as a constitutive limit of the full system describing
  the motion of an compressible linearly viscous fluid. To this end
  the starting system is written, using the Gibbs free energy, in the
  variables $\bv, \theta$ and $p$. The Oberbeck--Boussinesq system is
  then obtained as the thermal expansion coefficient $\alpha$
  and the isothermal compressibility coefficient $\beta$ tend to zero. 
\end{abstract} 
\maketitle
\section{Introduction} 

The well-known Oberbeck-Boussinesq \cite{Ober}, \cite{Bous} approximation was
designed as a simplified model for the thermo-mechanical response of
linear viscous fluids undergoing isochoric motions in isothermal
processes but not necessarily isochoric ones in non-isothermal
processes. Its roots stem from the end of the $19^{\mathrm{th}}$
century.  Nevertheless, its justification from the point of view of
continuum mechanics was quite recently given in 1996 by \cite{RRS}
(cf.~\cite{SV60}, \cite{HR91} and \cite{Mi62} for earlier
contributions).

From the mathematical point of view the expansion used in \cite{RRS}
is still formal. In \cite{krt1}, \cite{krt3} a rigorous justification
of simplified problems has been given. We refer the reader to
\cite{FN09}, \cite{FN09a} for a completely different approach in which
singular limits of the full system are discussed.

In this paper we use a new approach. Motivated by the
studies in \cite{bech03}, \cite{bech04} we obtain the
Oberbeck--Boussinesq approximation as a constitutive limit. Since this
limit is singular it depends on the way how it is achieved. To achieve
our result we re-write the full thermo-mechanical system with the help
of the Gibbs free energy in the variables $\bv, \theta$ and $p$. It is
important that the Gibbs free energy depends on two (dimensional)
parameters $a, b$, which tend to zero. Then we non-dimensionalize the
system and make assumptions on the behaviour of the thermal expansion
coefficient, the isothermal compressibility coefficient, the specific
heat coefficient at constant pressure and the "{}density''{} on the
non-dimensional versions of the paramaters $a, b$, which are denoted
by $A, B$. We show that these requirements can be fulfilled by an easy
example, which assumes that the density is linear in the pressure and
the temperatur.  Provided that weak solutions of the full
thermo-mechanical system satisfying a uniform estimate exists we show
that the limits satisfy the Oberbeck--Boussinesq
approximation. Moreover, if we assume that the approximation
parameters are fixed and sufficiently small we can also recover the
results obtained by a power series expansion in \cite{RRS}.

Let us finally introduce some notation: 
 
In what follows boldfaced minuscules always stand for vectors and 
vector valued functions whereas boldfaced capital letters represent 
tensor valued functions, i.e. $\bv=(v_1,v_2,v_3)^\top$, 
$\bT=(T_{kl})_{k,l=1}^3$; $\bT^\top:= (T_{lk})_{k,l=1}^3$.  
All quantities are considered at points $\bx=(x_1,x_2,x_3)^\top\in\R^3$ and 
at a certain time $t$.  We use the abbreviations 
$\partial_k:=\frac{\partial}{\partial x_k}$, 
$\partial_t:=\frac{\partial}{\partial t}$ (analogously for 
$\partial_{\theta}\dots$), 
$\partial^2_{\theta}:=\partial_{\theta}\partial_\theta$, 
$\di\bv:={\partial _1v_1}+{\partial_2 v_2}+{\partial_3 v_3}$ and 
$\n \rho\,:=({\partial_1 \rho\,},{\partial_2 \rho\,},{\partial_3 
\rho\,})^\top $.  
 The dot between two quantities denotes the corresponding scalar 
product, whereas the superposed dot is the usual material time 
derivative: \\ 
\hspace*{5mm}\parbox{14cm}{$\b{array}{rclrcl} 
\bv\cdot\bw&:=&\sum\limits_{k=1}^3 
v_kw_k,\qquad&\dot{\theta}&:=&\partial_t\theta+\bv\cdot\n\theta=\partial
_t\theta +
\sum\limits_{k=1}^3v_k \partial _k\theta,\\ 
 
\bT\cdot\bL&=&\sum\limits_{k,l=1}^3 T_{kl}L_{kl},& 
\dot{\bb}&:=& 
\partial_t\bb+[\nabla \bb]\bv=\partial_t\bb+\sum\limits_{k=1}^3v_k\partial_k\bb. 
\e{array}$}\\  
The trace of some tensor $\bD$ is denoted by tr $\bD$ and 
$\|\bD\|^2:= \bD\cdot \bD$. For the identity tensor we write 
$\bI$. 
%
\section{Derivation of the Approximation} 
\subsection{Governing Equations and Assumptions.}
The starting point for our analysis is the balance of mass, linear 
momentum and energy and the Second Law of Thermodynamics in the form 
of the Clausius-Duhem inequality:  
\ba   
  \dot{\rho\,}+\rho\,\di\bv&=&0,\nonumber
  \\ 
  \rho\,\dot{\bv}-\di\bT&=&\rho\,\bb, \label{2.1} 
  \\ 
  \rho\, \dot{e}-\di \bq &=&\bT\cdot\bD+\rho\, r,\nonumber
  \\
  \rho\,\dot{\eta}-\di\left(\frac\bq{\theta}\right)-\rho\,\frac r\theta&\ge& 0,\label{2.2}
\ea
where  $\rho\,$ denotes the density, $\bv$ the velocity field, 
$\bT$ the symmetric Cauchy stress tensor, $\bb$ the density of external body forces, 
$e$ the specific internal energy, $\bL$ the velocity gradient, $r$ the 
radiant heating, $\theta$ the temperature, $\eta$ the entropy and 
$\bq$  the heat flux vector.  

In the following we neglect radiant heating, i.e.~$r=0$, and assume
that  the body forces has a potential, i.e.~$\bb=\n f$. Moreover, we
restrict ourselves to the case of a compressible linearly viscous
fluid. Thus, we assume that
\begin{align}
  \begin{aligned}\label{eq:lin}
    \bT&=-p\, \bI +\lambda\,(\mbox{tr}\bD)\bI+2\mu\,\bD,
    \\
    \bq&= \kappa\, \nabla \theta,
  \end{aligned}
\end{align}
where $p $ is the pressure, $\lambda{}$, $\mu$ are the constant
viscosities and $\kappa{}$ is the constant thermal conductivity. 

If $\bv$, $\theta$ and $\rho$ are considered as independent variables
in \eqref{2.1}--\eqref{eq:lin} it is useful to introduce the Helmholtz
free energy $\psi$ through
\begin{align}\label{eq:helm}
  \psi(\rho,\theta):= e(\rho, \eta) -\theta\, \eta\,.
\end{align}
In this case we obtain from \eqref{eq:lin} that
\begin{align}
  \begin{gathered}\label{eq:rest-helm}
     \eta=-\pa _\theta \psi \,, \qquad  p =\rho^2 \,\pa _\rho \psi\,,
     \\
     \mu \ge 0\,, \qquad 3\lambda+2 \mu \ge 0\,, \qquad \kappa \ge 0\,,
  \end{gathered}
\end{align}
while \eqref{2.1} reads
\begin{align}
  \begin{aligned}\label{2.1h}
    \dot{\rho\,}+\rho\,\di\bv&=0\,,
    \\
    \rho\,\dot{\bv}-2\mu \,\di\bD -\lambda\, \n (\tr\bD) + \n
    \big ( \rho^2 \,\pa _\rho \psi\big )&=\rho\,\n f\,,
    \\
    -\rho\,\theta \Big ((\pa^2_\theta \psi )\,\dot{\theta} +
    (\pa^2_{\rho \theta } \psi )\,\dot{\rho}\Big) -\kappa\, \Delta
    \theta &=2\mu \, \abs{\bD}^2 +\lambda\, \abs{\tr \bD }^2 \,.
  \end{aligned}
\end{align}
However, for our purposes it is more convenient to view $\bv$,
$\theta$ and $p$ as independent variables in
\eqref{2.1}--\eqref{eq:lin}. To this end we introduce the Gibbs
free energy $\phi$ through
\begin{align}\label{eq:gibb}
  \phi(p,\theta):= \psi(\rho, \theta) + p\, \rho^{-1}\,.
\end{align}
In this situation we conclude from \eqref{eq:lin} that
\begin{align}
  \begin{gathered}\label{eq:rest-gibb}
     \eta=-\pa _\theta \phi \,, \qquad  \rho ^{-1}  =\pa _p \phi\,,
     \\
     \mu \ge 0\,, \qquad 3\lambda+2 \mu \ge 0\,, \qquad \kappa \ge 0\,.
  \end{gathered}
\end{align}
The system \eqref{2.1} now reads
\begin{align}
  \begin{aligned}\label{2.1g}
    \big(\pa _p\phi \big)^{-1} \Big ( (\pa^2_{\theta p} \phi
    )\,\dot{\theta} +(\pa^2_p \phi )\,\dot{p} \Big )&= \di\bv\,,
    \\
    \big(\pa _p\phi \big)^{-1} \dot{\bv}-2\mu \,\di\bD -\lambda\, \n
    (\tr\bD) + \n p&=\big(\pa _p\phi \big)^{-1} \n f\,,
    \\
    -\theta\, \big(\pa _p\phi \big)^{-1} \Big ((\pa^2_\theta \phi
    )\,\dot{\theta} + (\pa^2_{\theta p} \phi )\,\dot{p} \Big) -\kappa\,
    \Delta \theta &=2\mu \, \abs{\bD}^2 +\lambda\, \abs{\tr \bD }^2
    \,.
  \end{aligned}
\end{align}
Equation \eqref{2.1g}$_1$ nicely reflects the fact that changes in
volume are induced by changes in temperature and changes in pressure.
In fact, if we introduce the thermal expansion coefficient $\alpha$
and the isothermal compressibility coefficient $\beta$ through
\begin{align}
  \begin{aligned}\label{eq:coeff}
    \alpha(p, \theta)&:= \big(\pa _p\phi \big)^{-1} \pa^2_{\theta p}
    \phi\,,
    \\
    \beta(p, \theta)&:= -\big(\pa _p\phi \big)^{-1} \pa^2_{p}
    \phi\,,
  \end{aligned}
\end{align}
we can re-write \eqref{2.1g}$_1$ as 
\begin{align}\label{eq:mass}
    \alpha\, \dot{\theta} - \beta \,\dot{p} &= \di\bv\,.
\end{align}
It is well known that for many fluids the thermal expansion
coefficient~$\alpha$ is small ($\alpha \approx (10^{-4} - 10^{-3}) {\rm K}^{-1}$) and
the isothermal compressibility coefficient $\beta$ is even smaller
($\beta \approx (10^{-11} - 10^{-10}) {\rm Pa}^{-1}$). Finally, it is convenient to
introduce the specific heat coefficient at constant pressure through
\begin{align}\label{def:cp}
  c_p(p, \theta):= -\theta\, \pa^2_\theta \phi \,.
\end{align}
With this notation we can re-write \eqref{2.1g}$_3$ as 
\begin{align}
   c_p  \, \big(\pa _p\phi \big)^{-1} \,\dot{\theta} - \alpha\, \theta
   \, \dot{p}  -\kappa\,
    \Delta \theta &=2\mu \, \abs{\bD}^2 +\lambda\, \abs{\tr \bD }^2
    \,.
\end{align}
\begin{remark}{\rm
  Note that in \cite{RRS} special fluids that can sustain isochoric
  motions in isothermal processes have been considered. This formally
  corresponds to neglecting $\beta$ in \eqref{eq:mass}. For such
  fluids the Oberbeck--Boussinesq approximation has been formally
  derived in \cite{RRS} with the help of a power series expansion. In
  \cite{krt1}, \cite{krt3} a rigorous mathematical justification of a
  simplified modell has been carried out. However, the mathematical
  justification starting with the full system from \cite{RRS} is still
  lacking. One of the difficulties is the lack of appropriate apriori
  estimates, which is related to the fact that $\beta$ has been
  neglected in \eqref{eq:mass}.}
\end{remark}

In \cite{bech03}, \cite{bech04} thermal expansion models have been
considered as a constitutive limit for free energies. In these papers
the predictions of the compressible theory has been compared to the
prediction of different limiting theories. However, these limiting
theories are from the mathematical point of view singular
limits. Thus, the way how the limit is achieved is important and
different ways can result in different limiting systems.

In the present paper we combine ideas from \cite{RRS},
\cite{krt1} and \cite{bech03}, \cite{bech04} in order to derive
the Oberbeck--Boussinesq approximation as a constitutive limit. To
this end we consider the system \eqref{2.1g} for a family of Gibbs
free energies depending on (dimensional) parameters $a,b>0$, i.e.
\begin{align*}
  \phi(p,\theta)= \phi^{a,b}(p, \theta)\,.
\end{align*}
The dependence on the parameters $a, b$ will be suppressed in the
notation in most cases.  Under certain assumptions we consider the
limit as $a,b$ tend to zero.  Before that we render the system
\eqref{2.1g} non-dimensional by introducing dimensionless variables
\b{eqnarray*} 
  \overline{\bx}:= \frac \bx L, \quad \overline{t}:=\frac t {T}, \quad
  \overline{\bv}:= \frac\bv V, \quad \overline{p}:=\frac p{\pi}, \quad
  \overline{\theta}:=\frac{\theta}{\vartheta}-\theta _r, \quad
  \overline{f}:=\frac{f}{gL},
\e{eqnarray*} 
where $L, T, V, \pi$ and $\theta_r, \vartheta$ are typical length,
time, velocity, pressure and temperatures, while $g$ is the
gravitational constant.\footnote{ This form of the
  non-dimensionalization is motivated by a typical situation when the
  Oberbeck--Boussinesq approximation is used, namely a fluid layer in
  a gravitational field with a certain difference between the
  temperatur $\theta ^t$ at the top and $\theta ^b$ at the botton. In
  this case we would choose $\vartheta=\theta ^b - \theta^t$ and
  ${\theta_r=2^{-1}\vartheta ^{-1}(\theta ^t + \theta^b)}$.} Moreover,
let $\phi_0$ be a typical value for the Gibbs free energy. We will not
choose $V$, $L$ and $T$ independent of one another but use the
relation $V\,T=L$. Note that the constants $\mu, \lambda, \kappa,
\theta_r$ and $g$ are assumed to be independent of the parameters
$a,b$, while $V, T, L, \vartheta, \pi$ and $ \phi_0$ may depend on the
parameters $a, b$. We will make this dependence explicit if necessary.

The definition of the non-dimensional analogues of $a$ and $b$ is
motivated by the following observation: We introduce non-dimensional a
thermal expansion coefficient $\ov{\alpha}$ and the isothermal
compressibility coefficient $\ov{\beta}$ by
\begin{gather*}
  \ov{\alpha}(\ov p, \ov \theta):= \frac {\alpha
    (p,\theta)}{\alpha_0}\,,\qquad \ov{\beta}(\ov p, \ov \theta):=
  \frac {\beta (p,\theta)}{\beta_0}\,,
\end{gather*}
where $\alpha_0$ and $\beta_0$ are typical values for for the thermal
expansion coefficient $\alpha$ and the isothermal compressibility
coefficient $\beta$. The non-dimensional version of \eqref{eq:mass}
then reads 
\begin{align}\label{eq:mass-n}
    \alpha_0 \vartheta\, \ov \alpha\, \dot{\ov \theta} - \beta_0\, \pi\,
    \ov \beta \,\dot{\ov p} &= \ov \di \ov \bv\,,
\end{align}
where the superposed dot now stands for the non-dimensional material
time derivative. In typical applications we see that
\begin{gather}\label{ex-num}
  \alpha_0 \vartheta  \approx 10^{-3} - 10^{-2}
  \,, \qquad 
  \beta_0 \pi  \approx 10^{-6} - 10^{-5}\,.
\end{gather}
Since $\alpha_0 \vartheta$ and $\beta_0 \pi$ are non-dimensional
numbers (depending on the parameters $a, b$) we roughly want that the
non-dimensional analogue $A $ of $a$ behaves as $\alpha_0 \,
\vartheta$ and the non-dimensional analogue $B $ of $b$ behaves as
$ \beta_0\, \pi$. This is made precise in the following way: We set 
\begin{align}\label{def:hab}
  h^{a,b}(\ovp,\ovt):=\frac {\phi^{a,b}(p,\theta)}{\phi_0^{a,b}}
\end{align}
and define the non-dimensional parameters $A, B$ through
\begin{align}
  \begin{aligned}\label{def:AB}
    A=A(a,b)&:= \big(\pa _{\ovp}\, h^{a,b}(1,1) \big)^{-1} \pa^2_{\ovt
      \ovp}\, h^{a,b}(1,1)\,,
    \\
    B=B(a,b)&:= -\big(\pa _{\ovp}\, h^{a,b}(1,1) \big)^{-1} \pa^2_{\ovp}\,
    h^{a,b}(1,1)\,.
  \end{aligned}
\end{align}
Moreover, we assume that the mapping $(a,b) \mapsto (A,B)$ is
invertible and that $(a,b) \to (0,0)$ implies $(A,B)\to
(0,0)$. Finally, set
\begin{align}\label{def:pAB}
  \ov{\phi}^{A,B}(\ovp,\ovt):= \frac {\phi^{a(A,B),b(A,B)}
    (p,\theta)}{\phi_0^{a(A,B),b(A,B)}}\,. 
\end{align}
The system \eqref{2.1g} for the non-dimensional quantities and
differential operators then becomes (we skip all bars for
convenience):
\begin{align}
  \begin{gathered}\label{2.1gn}
    \big(\pa _p\phiab \big)^{-1} \Big ( (\pa^2_{\theta p} \phiab
    )\,\dot{\theta} +(\pa^2_p \phiab )\,\dot{p} \Big )= \di\bv\,,
    \\[2mm]
    \begin{aligned}
      &\big(\pa _p\phiab \big)^{-1} \dot{\bv}- \frac
      {\phi_0}{\pi\,L\,V} \Big (2\mu \,\di\bD +\lambda\, \n
      (\tr\bD)\Big ) + \frac {\phi_0}{V^2}\,\n p
      \\
      &= \frac{g\,L}{V^2}\big(\pa _p\phiab \big)^{-1} \n f\,,
    \end{aligned}
    \\[2mm]
    \begin{aligned}
      &-\big (\theta +\theta_r \big)\, \big(\pa
      _p\phiab \big)^{-1} \Big ((\pa^2_\theta \phiab )\,\dot{\theta} +
      (\pa^2_{\theta p} \phiab )\,\dot{p} \Big) -\frac
      {\kappa\,\vartheta}{\pi\, L\, V} \,\Delta \theta
      \\
      &=\frac{V}{\pi\, L}\Big(2 \mu \abs{\bD}^2 +\lambda\, \abs{\tr
        \bD }^2 \Big )\,,
    \end{aligned}
  \end{gathered}
\end{align}
where we used the notation $\phi_0:=\phi_0^{a(A,B),b(A,B)}$. Of
course, for the quantities $V, T, L, \vartheta$ and $ \pi$ the same
convention is used, e.g.~$V=V^{a(A,B),b(A,B)}$.

Apriori there is no obvious representative velocity $V$ in natural
convection processes. As already observed in \cite{Ch61} such
processes are reflected by the assumption $V^2 \approx 
{g\,L}\,\alpha_0\,\vartheta$. This is translated in our situation by 
\begin{align}\label{def:V}
  V^2 := A\, g\,L\,.
\end{align}
From \eqref{2.1gn}$_2$ follows that non-trivial body forces are only possible
if ${\phi_0 \approx g\, L}$. Consequently, we set 
\begin{align}\label{def:gamma}
  \gamma:= \frac {\phi_0}{g\,L}\,,
\end{align}
and require that $\gamma=O(1)$ as $A, B$ tend to zero. We also
introduce the Reynolds numbers $\Re_\mu$ and $\Re_\lambda$ as well as
the Prandtl number $\Pr$ by setting
\begin{align}\label{def:rp}
  \Re_\mu := \frac{\pi\, V\,L}{2\,\mu\, \phi_0}\,, \qquad \Re_\lambda :=
  \frac{\pi\, V\,L}{\lambda\, \phi_0}\,, \qquad \Pr :=
  \frac{\phi_0\,\mu }{\vartheta \, \kappa}\,,
\end{align}
and require that $\Re_\mu =O(1)$, $\Re_\lambda =O(1)$, $\Pr =O(1)$
as  $A, B$ tend to zero. Note that all these requirements can be
fulfilled simultaneously, e.g.~if
\begin{gather}\label{eq:behavA}
  V\approx A^{\frac 13}\,, \quad   L\approx A^{\frac {-1}3}\,, \quad
  \phi_0\approx A^{\frac {-1}3}\,, \quad \pi\approx A^{\frac {-1}3}\,, \quad
  \vartheta \approx A^{\frac {-1}3}\,,
\end{gather}
then all above requirements are satisfied.

Using the above notation we can re-write \eqref{2.1gn} as 
\begin{align}
  \begin{gathered}\label{eq:final}
    \big(\pa _p\phiab \big)^{-1} \Big ( (\pa^2_{\theta p} \phiab
    )\,\dot{\theta} +(\pa^2_p \phiab )\,\dot{p} \Big )= \di\bv\,,
    \\[2mm]
    \begin{aligned}
      &\big(\pa _p\phiab \big)^{-1} \dot{\bv}- \frac
      {1}{\Re _\mu} \,\di\bD -\frac 1 {\Re_\lambda}\, \n
      (\tr \bD) + \frac {\gamma}{A}\,\n p
      \\
      &= \frac{1}{A}\big(\pa _p\phiab \big)^{-1} \n f\,,
    \end{aligned}
    \\[2mm]
    \begin{aligned}
      &-\big (\theta +\theta_r \big)\, \big(\pa
      _p\phiab \big)^{-1} \Big ((\pa^2_\theta \phiab )\,\dot{\theta} +
      (\pa^2_{\theta p} \phiab )\,\dot{p} \Big) -\frac {1}{\Pr\,
        \Re_\mu} \,\Delta \theta
      \\
      &=A \,\Big(\frac{1}{\gamma\,\Re_\mu} \abs{\bD}^2 +
      \frac{1}{\gamma\,\Re_\lambda}\, \abs{\tr \bD }^2 \Big )\,.
    \end{aligned}
  \end{gathered}
\end{align}
Concerning the behaviour with respect to $A, B$ of the remaining
quantities "thermal expansion coefficient'', "isothermal
compressibility coefficient'', "specific heat at constant pressure''
and "density'' in \eqref{eq:final} we make the following assumptions:
There exists constants $c_0>0, k_i^{A,B}$, $i=1,2$, and functions
$\alpha_i^{A,B}(p,\theta)$, $\beta_i^{A,B}(p,\theta)$,
$c_i^{A,B}(p,\theta)$, $\rho_r^{A,B}(p,\theta)$, $i=1,2$, such that
\begin{align}
  \begin{aligned}\label{ass:AB}
    \frac {\pa^2_{\theta p} \phiab(p,\theta)}{\pa
      _p\phiab(p,\theta)}&= A\big ( 1 +A\, \alpha^{A,B}_1(p,\theta)
    +B\, \alpha^{A,B}_2(p,\theta) \big)\,,
    \\[2mm]
    \frac {\pa^2_{p} \phiab(p,\theta)}{\pa _p\phiab(p,\theta)}&=
    -B\big ( 1 +A\, \beta^{A,B}_1(p,\theta) +B\,
    \beta^{A,B}_2(p,\theta) \big)\,,
    \\[2mm]
    \frac 1 {\pa _p\phiab (p,\theta)} &= \big (1 - A\,\big (\theta +
    {\theta_r} ) + B\, p + \rho^{A,B}_r(p,\theta) \big ) k_1^{A,B}\,,
    \\[2mm]
    -\big (\theta +\theta_r \big)\, \pa^2_\theta \phiab(p,\theta ) &=
    \big ( c_0 \!+\!A^2\,c^{A,B}_1(p,\theta) +B^2\,
    c^{A,B}_2(p,\theta) \big ) k_2^{A,B}\,,
  \end{aligned}
\end{align}
where we require that 
\begin{gather}
  \label{eq:ki}
  \lim_{(A,B)\to (0,0)} k_i^{A,B}=1\,,
\end{gather}
and locally uniformly in $p, \theta$
\begin{align}
  \begin{gathered}\label{eq:AB1}
    \alpha^{A,B}_i(p,\theta) = O(1)\,, \quad \beta^{A,B}_i(p,\theta) =
    O(1)\,, \quad c^{A,B}_i(p,\theta) = O(1)\,, 
    \\
    \rho^{A,B}_r(p,\theta) = O(A^2) + O(B^2) + O(A\,B)\,
  \end{gathered}
\end{align}
with $ i=1,2$, as $A, B $ tend to zero.

Let us illustrate the above procedure and assumptions by the following example:
\begin{example}{\rm
  We assume that 
  \begin{align*}
    \rho^{a,b}(p,\theta)= \big(\pa _p\phi^{a,b}(p,\theta) \big)^{-1} =
    \rho_0\big (1 +b\, p -a\, \theta\big )\,,
  \end{align*}
  i.e.~the density depends linearly on the pressure and the
  temperature. This constitutive equation is in dimensional form and
  the constants $\rho_0, a, b$ are assumed to be positive. A possible
  Gibbs free energy compatible with such a behaviour is given by
  \begin{align*}
    \phi^{a,b}(p,\theta ) &= b^{-1}\rho^{-1}_0 \big (\ln (1
    +b\, p -a\, \theta) - \ln (1 -a\, \theta)\big ) - c_0 \,
    \theta \, (\ln \theta -1) \,,
  \end{align*}
  where $c_0>0$. From the definition in \eqref{def:AB} we easily compute
  \begin{align}
    \begin{aligned}\label{eq:A-B}
      A=A(a,b)= a\, \vartheta \big (1 +b\, \pi -a\,\vartheta (1 +
      \theta_r) \big)^{-1}\,,
      \\
      B=B(a,b)= b\, \pi \big (1+b\, \pi -a\,\vartheta (1+
      \theta_r) \big)^{-1}\,.
    \end{aligned}
  \end{align}
  We set
  \begin{gather}
    \label{def:pi-theta}
    \vartheta:= \vartheta_0 \big ( a\, \vartheta _0\big )^{\frac
      {-1}4}\,, \quad \pi:= \pi_0 \big ( a\, \vartheta _0\big )^{\frac
      {-1}4}\,, \quad \phi_0^{a,b}:= \phi ^{a,b} \Big (\pi, \vartheta
    \frac {b\, \pi_0}{ a\, \vartheta _0 } \Big )\,,
  \end{gather}
  where $\vartheta _0$ and $\pi_0$ are positive constants independent
  of $a,b$.  Moreover, we require that 
  \begin{align}\label{eq:add}
    \lim _{(a,b)\to (0,0)} \frac {b\, \pi_0}{ (a\, \vartheta _0 )^{5/4}} =0\,.
  \end{align}
  Note that in typical situations this requirement is fulfilled
  (cf.~\eqref{ex-num}). Moreover, this requirement ensures that
  $(A,B)\to (0,0)$ if $(a,b)\to (0,0)$.  The system of equations
  \eqref{eq:A-B} for $a, b$ is solvable and we compute
  \begin{align}
    \begin{aligned}
      a&= \frac 1 {\vartheta_0} \, \left( \frac {A } { 1 + A (1 +
          {\theta _r} ) -B}\right)^{ 4/3}=: \frac 1 {\vartheta_0}
      \, \left( \frac {A } {x_{A,B}}\right)^{ 4/3}\,,
      \\
      b&= \frac 1 {\pi_0} \, \frac {B\, A^{1/3}}  {x_{A,B}^{ 4/3}} \,.
    \end{aligned}
  \end{align}
  Moreover, straightforward manipulations show (with skipped bars)
  \begin{align}
    \begin{aligned}\label{ass:AB-ex}
      \frac {\pa^2_{\theta p} \phiab(p,\theta)}{\pa
        _p\phiab(p,\theta)}&= A\Big ( 1 - \frac { B\, (p-1)-A\,(\theta
        -1) }{1 +B\, (p-1)-A\,(\theta-1 ) } \Big)\,,
      \\[2mm]
      \frac {\pa^2_{p} \phiab(p,\theta)}{\pa _p\phiab(p,\theta)}&=-B
      \Big ( 1 - \frac { B\, (p-1)-A\,(\theta -1) }{1 +B\,
        (p-1)-A\,(\theta-1 ) } \Big)\,,
      \\[2mm]
      \frac 1 {\pa _p\phiab (p,\theta)} &= \big (1 - A\,(\theta
      +\theta_r ) + B\, p + \rho^{A,B}_r(p,\theta)\big ) k_1^{A,B}\,,
      \\[2mm]
      -\big (\theta +\theta_r \big)\, \pa^2_\theta \phiab(p,\theta )
      &= \Big ( \frac {c_0 \vartheta _0 \rho_0}{\pi_0}
      +A^2\,c^{A,B}_1(p,\theta) \Big ) k_2^{A,B}\,,
    \end{aligned}
  \end{align}
  where
  \begin{align*}
    \rho^{A,B}_r(p,\theta)&= -AB \Big ((p-1) \frac {1+\theta
      _r}{x_{A,B}}+\frac {\theta -1}{x_{A,B}}\Big ) +A^2 (\theta-1)
    \frac {1+\theta _r}{x_{A,B}}+B^2\frac {p -1}{x_{A,B}}\,,
    \\[2mm]
    k_1^{A,B}&=\frac {x_{A,B}}{B} \Big ( \ln \big ( 1+ \frac
    B{x_{A,B}} -\frac B{x_{A,B}^{13/12}} \big ) - \ln \big ( 1 -\frac
    B{x_{A,B}^{13/12}} \big ) \Big )
    \\
    &\quad -A^{1/3} \frac {c_0 \rho _0 \vartheta_0}{\pi _0} \frac 1 {x_{A,B}^{1/12}}\frac
    {B}{A^{4/3}}\Big (\ln \big ( \vartheta _0 x_{A,B}^{1/4}\frac
    {B}{A^{4/3}}\big ) -1\Big )\,,
    \\
    c_1^{A,B}(p,\theta)&= \frac {- x_{A,B} \,p\,(\theta +\theta _r)\big (2+B(p-2)-2A(\theta
      -1)\big )}{\big (1 +B\, (p-1)-A\,(\theta-1 ) \big)^2 \big(1 -B
      -A\,(\theta-1 ) \big )^2}\,,
    \\
    k_2^{A,B}&=\big (k_1^{A,B}\big )^{-1} \,.
  \end{align*}
  Note that due to \eqref{eq:add} we have $B\, A^{-4/3} \to 0$ if
  $A,B$ tend to zero. Thus, our example fulfills the requirements
  \eqref{ass:AB}--\eqref{eq:AB1} as well as the requirements for
  \eqref{def:gamma} and \eqref{def:rp}.}
\end{example}

\subsection{Constitutive limit.} 
Now we want to show that we are able to obtain the
Oberbeck--Boussinesq approximation as a constitutive limit of
\eqref{eq:final} as $A, B$ tend to zero. We assume that the
requirements \eqref{ass:AB}--\eqref{eq:AB1} as well as the
requirements for \eqref{def:gamma} and \eqref{def:rp} are
fulfilled. We consider solutions $\bv =\bv^{A,B}$,
$\theta =\theta^{A,B}$ and $p=p^{A,B}$ of \eqref{eq:final} in the
following sense:\\
Let $\Omega \subset \R^3$ be a given sufficiently smooth bounded
domain and ${I=(0,T)}$, $T>0$, be a given time intervall. We set
$Q=I\times \Omega$ and assume that ${\n f \in L^2(Q)}$ and appropriate
boundary and initial conditions for $\bv$ and $\theta $ are given. The system
\eqref{eq:final} is satisfied in the sense of distributions over $Q$
and $\bv$, $\theta$ and $p$ are uniformly bounded in
$W^{1,2}(Q)\cap L^\infty(Q)$ with respect to $A$, $B$. Finally, we
assume that $B=o(A)$.

From our assumptions follows that there exists $\bv$, $\theta, p \in
W^{1,2}(Q)$ such that
\begin{align*}
  \bv^{A,B} &\rightharpoonup \bv \qquad \textrm{weakly in }W^{1,2}(Q),
  \\
  \theta^{A,B} &\rightharpoonup \theta \,\qquad \textrm{weakly in }W^{1,2}(Q),
  \\
  p^{A,B} &\rightharpoonup p \,\qquad \textrm{weakly in }W^{1,2}(Q)
\end{align*}
and for all $q\in [1,\infty)$
\begin{align*}
  \bv^{A,B} &\to \bv \qquad \textrm{strongly in }L^{q}(Q),
  \\
  \theta^{A,B} &\to \theta \,\qquad \textrm{strongly in }L^{q}(Q),
  \\
  p^{A,B} &\to p \,\qquad \textrm{strongly in }L^{q}(Q).
\end{align*}
These convergences and the assumptions \eqref{ass:AB}--\eqref{eq:AB1}
as well as the requirements for \eqref{def:gamma} and \eqref{def:rp}
 immediately imply that the limit as $A, B$ tend to zero in
\eqref{eq:final}$_{1,3}$ yield for all $\bphi, \psi \in C^\infty_0(Q)$
\begin{align}
  \label{eq:11}
  \begin{split}
    \int_Q \di \bv \cdot \bphi \, dx\, dt&=0,
    \\
    \int_Q c_0\,\pa _t \theta \,\psi + c_0\,\bv\cdot \n \theta \, \psi + \frac
    1{\Pr \Re_\mu} \n \theta \cdot \n \psi \, dx \, dt &=0.
  \end{split}
\end{align}
To treat \eqref{eq:final}$_2$ we add $k_1^{A,B}\n f$ on both sides and
obtain for all $\bphi \in C^\infty_0(Q)$
\begin{align*}
  &\int _Q k_1^{A,B} \big (1 - A\,\big (\theta^{A,B} +
  {\theta_r} ) + B\, p^{A,B} + \rho^{A,B}_r(p^{A,B},\theta^{A,B}) \big
  ) \times 
  \\
  &\quad \;\;\times \big (\partial _t \bv^{A,B} + [\n \bv^{A,B}]\bv
  ^{A,B}\big )\cdot \bphi \, dx\, dt
  \\
  &+\int_Q \frac 1 {\Re_\mu} \bD \bv ^{A,B} \cdot \bD \bphi +  \frac 1
    {\Re_\lambda} \di \bv ^{A,B} \, \di \bphi \, dx \, dt
  \\
  &-\int_Q \Big ( \frac {\gamma \, p^{A,B}}{A} +k_1^{A,B}f (1-\theta
    _r) - \frac {k_1^{A,B}}{A} f\Big)\, \di \bphi \, dx \, dt
  \\
  &= \int_Q k_1^{A,B}\Big ( \big (1-\theta ^{A,B}\big ) +  \frac BA \, p^{A,B}
    + \frac 1 A\rho^{A,B}_r(p^{A,B},\theta^{A,B})  \Big )\n f \cdot
    \bphi \,dx \, dt.
\end{align*}
From this we deduce for $A, B$ tending to zero for all $\bphi \in
C^\infty _0(Q)$ with $\di \bphi =0$
\begin{align}
  \begin{split}
    &\int _Q \big (\partial _t \bv + [\n \bv]\bv \big )\cdot \bphi \,
    dx\, dt + \frac 1 {\Re_\mu} \int_Q \bD \bv \cdot \bD \bphi \, dx
    \, dt
    \\
    &= \int_Q \big (1-\theta \big )\n f \cdot \bphi \,dx \, dt.
  \end{split}\label{eq:12}
\end{align}
The system \eqref{eq:11}, \eqref{eq:12} is the weak formulation of the
celebrated Oberbeck--Boussinesq approximation.

\subsection{Formal expansion.} 
Our approach enables us also to recover the results in \cite{RRS} and
\cite{krt1} for dissipation number $\textrm {Di}=0$. To this end  we
assume that the parameters $A, B$ are fixed but sufficiently
small. Moreover, we assume that $A$ is smaller than $B$ in the sense
that $A^2 \le B \le A$. This requirement is fulfilled in typical
applications. In this situation we formally expand
the non-dimensional quantities $\bv$, $\theta$ and $p$ in
\eqref{eq:final} into power series with respect to the perturbation
parameters $A, B$ in the form
\begin{align*}
\bv=\!\!\sum_{m,k=0}^\infty \!\!A^{m}B^k\bv_{m,k},\;\;
\theta=\!\!\sum_{m,k=0}^\infty \!\! A^m B^{k}\theta_{m,k},\;\;
p=\!\!\sum_{m,k=0}^\infty \!\! A^m B^{k}p_{m,k}. 
\end{align*}
We attach the boundary conditions of the quantities to the zero order
term and set the boundary conditions to be zero for the higher order
terms. Similarly we expand the quantities $c_0, k_i^{A,B}$, $i=1,2$,
and functions $\alpha_i^{A,B}(p,\theta)$, $\beta_i^{A,B}(p,\theta)$,
$c_i^{A,B}(p,\theta)$, $\rho_r^{A,B}(p,\theta)$, $i=1,2$, where we
also insert the series for $p$ and $\theta$ in the arguments.  We
replace all these quantities in the system \eqref{eq:final} by the
corresponding power series.  At level $(0,0)$ we see from
(\ref{eq:final}) that\footnote{To simplify the notation we use the
  superposed dot only at the level $(0,0)$.}
\begin{align*}
  \di \bv_{0,0}&=0,
  \\
  \gamma \n p_{0,0}&=\n f, 
  \\
  c_0 \dot{\theta}_{0,0}  -\frac 1{\Pr \Re _\mu}\Delta\theta_{0,0}&=0.
\end{align*}
At level $(1,0)$ we see from (\ref{eq:final})$_2$ that
\begin{align*}
  \dot{\bv}_{0,0}  -\frac 1{\Re _\mu}\di \bD \bv_{0,0} + \gamma \n
  p_{1,0}&= -(\theta _{0,0}- \theta _r)\n f + k_{1; 1,0}\n f .
\end{align*}
Thus, setting $\bv:=\bv_{0,0}$, $\theta:= \theta_{0,0}$ and
$p:= \gamma p_{0,0}+A ( \gamma p_{1,0} - \theta _r f -k_{1;1,0} f)$ we
obtain 
\begin{align}
  \label{eq:theend}
  \begin{split}
      \di \bv&=0,
      \\
      A\Big (\dot{\bv}  -\frac 1{\Re _\mu}\di \bD \bv \Big ) + \n
      p&= (1- A\,\theta )\n f,
      \\
      c_0 \dot{\theta}  -\frac 1{\Pr \Re _\mu}\Delta\theta&=0.
  \end{split}
\end{align}
This system is  the same as the one obtained in \cite{RRS} and
\cite{krt1} for $\textrm {Di}=0$.  The approximation combines different
levels in the temperature and the velocity equation and the velocity
has zero divergence.  

\smallskip

\section*{Acknowledgement.} 
 Y. K. was partly supported
by JSPS KAKENHI Grant Number 24340028, 22244009, 24224003, 15K13449. 
M.R.~would like to thank for the financial support and the hospitality
during several visits at the Kyushu University. 



\b{thebibliography}{99} 
\small 
\bibitem{bech03}
Bechtel,  S.~E., Forest, M.~G., Rooney, F.~J., Wang,Q.: Thermal
  expansion models of viscous fluids based on limits of free energy, Physics
  of Fluids {\bf 15}, 2681 (2003)
\bibitem{bech04}
Bechtel, S.~E., Cai, M., Rooney, F.~J., Wang, Q.: Investigation of
  simplified thermal expansion models for compressible Newtonian fluids applied
  to nonisothermal plane Couette and Poiseuille flows, Physics of Fluids {\bf
  16},  3955 (2004)

\bibitem{Bous} 
Boussinesq, J.: Th\'eorie Analytique de la Chaleur. Paris: Gauthier-Villars, 1903 

\bibitem{Ch61}
Chandrasekhar,S.:  The International Series of Monographs on Physics,
  Clarendon Press, Oxford, 1961.

\bibitem{krt1}
Kagei, Y., R{\r {u}}{\v{z}}i{\v{c}}ka, M., Th{\"a}ter, G.: Natural
  Convection with Dissipative Heating, Comm. Math. Phys. {\bf 214},
  287 (2000)

\bibitem{krt3}
Kagei, Y., R{\r {u}}{\v{z}}i{\v{c}}ka, M., Th{\"a}ter, G.: A Limit
Problem in Natural Convection, NoDEA {\bf 13}, 447 (2006)

\bibitem{FN09a}
Feireisl, E., Novotn{\'y}, A.: The {O}berbeck-{B}oussinesq
  approximation as a singular limit of the full {N}avier-{S}tokes-{F}ourier
  system, J. Math. Fluid Mech. {\bf 11}, 274 (2009)

\bibitem{FN09}
Feireisl, E., Novotn{\'y}, A.: Advances in Mathematical Fluid Mechanics, Birkh\"auser Verlag,
  Basel, 2009

\bibitem{HR91}
Hills, R.~N., Roberts, P.~H.: On the motion of a fluid that is
  incompressible in a generalized sense and its relationship to the
  {B}oussinesq approximation, Stability Appl. Anal. Contin. Media {\bf
    1}, 205 (1991)

\bibitem{Mi62}
Mihaljan, J.~M.: A rigorous exposition of the {B}oussinesq
  approximations applicable to a thin layer of fluid,
  Astrophys. J. {\bf 136}, 1126  (1962)

\bibitem{Ober} 
Oberbeck, A.: \"Uber die W\"armeleitung der Fl\"ussigkeiten bei der Ber\"ucksichtigung der Str\"omungen infolge von Temperaturdifferenzen. 
Annalen der Physik und Chemie {\bf 7},  271 (1879);\\
\"Uber die Bewegungserscheinungen der Atmosph\"are. Sitz. Ber. K. Preuss. Akad. Miss. 383 and 1120 (1888) 
\bibitem{RRS} 
Rajagopal,  K. R.,  \rose, M.,  Srinivasa, A. R.: On the Oberbeck-Boussinesq Approximation. Math. Models Methods 
Appl. Sci. {\bf 6}, 1157--1167 (1996) 
\bibitem{SV60}
Spiegel, E.~A., Veronis, G.: On the {B}oussinesq approximation for
  a compressible fluid, Astrophys. J. {\bf 131}, 442 (1960)

\e{thebibliography}
\e{document}